\documentclass[iop]{emulateapj}
\usepackage{lscape}
\usepackage{rotating}

\slugcomment{Accepted for publication on The Astrophysical
Journal, 9 December 2012} \shortauthors{Tortora et al.}
\shorttitle{An inventory of the Initial Mass Function}

\usepackage{color}

\usepackage{natbib}

\usepackage{aas_macros}

\def\sigs{\mbox{$\sigma_\star$}}

\def\Re{\mbox{$R_{\rm e}$}}

\def\Msun{\mbox{$M_\odot$}}

\def\dimf{\mbox{$\delta_{\rm IMF}$}}
\def\Yst{\mbox{$\Upsilon_\star$}}

\def\YstMW{\mbox{$\Upsilon_{\star, \rm MW}$}}

\def\Ydyn{\mbox{$\Upsilon_{\rm dyn}$}}

\def\mst{\mbox{$M_{\star}$}}

\def\Mvir{\mbox{$M_{\rm vir}$}}
\def\cvir{\mbox{$c_{\rm vir}$}}

\def\fdm{\mbox{$f_{\rm DM}$}}

\def\lsim{\mathrel{\rlap{\lower3.5pt\hbox{\hskip0.5pt$\sim$}}
    \raise0.5pt\hbox{$<$}}}                
\def\gsim{~\rlap{$>$}{\lower 1.0ex\hbox{$\sim$}}}
\def\Rap{\mbox{$R_{\rm Ap}$}}

\def\sigAp{\mbox{$\sigma_{\rm Ap}$}}
\def\sige{\mbox{$\sigma_{\rm e}$}}

\newcommand{\appropto}{\mathrel{\vcenter{
  \offinterlineskip\halign{\hfil$##$\cr
    \propto\cr\noalign{\kern2pt}\sim\cr\noalign{\kern-2pt}}}}}

\begin{document}

\title{An inventory of the stellar initial mass function in early-type galaxies}

\author{C.~Tortora\altaffilmark{1}\footnote{\texttt ctortora@physik.uzh.ch},
A.J.~Romanowsky\altaffilmark{2,3}, and
N.R.~Napolitano\altaffilmark{4}}

\affil{
\altaffilmark{1}Universit$\rm \ddot{a}$t Z$\rm \ddot{u}$rich,
Institut f$\rm \ddot{u}$r Theoretische Physik, Winterthurerstrasse
190, CH-8057, Z$\rm \ddot{u}$rich, Switzerland\email{\texttt
ctortora@physik.uzh.ch}\\
\altaffilmark{2}University of California Observatories, 1156 High Street, Santa
Cruz, CA 95064, USA\\
\altaffilmark{3} Department of Physics and Astronomy, San Jos\'e
State University, One Washington Square, San Jose, CA 95192, USA\\
\altaffilmark{4} INAF -- Osservatorio Astronomico di Capodimonte,
Salita Moiariello, 16, 80131 - Napoli, Italy }

\begin{abstract}
Given a flurry of recent claims for systematic variations in the
stellar initial mass function (IMF), we carry out the first
inventory of the observational evidence using different
approaches. This includes literature results, as well as our own
new findings from combined stellar-populations synthesis (SPS) and
Jeans dynamical analyses of data on $\sim$~4500 early-type
galaxies (ETGs) from the SPIDER project. We focus on the
mass-to-light ratio mismatch relative to the Milky Way IMF, \dimf,
correlated against the central stellar velocity dispersion, \sigs.
We find a strong correlation between \dimf\ and \sigs, for a wide
set of dark matter (DM) model profiles. These results are robust
if a uniform halo response to baryons is adopted across the
sample. The overall normalization of \dimf, and the detailed DM
profile, are less certain, but the data are consistent with
standard cold-DM halos, and a central DM fraction that is roughly
constant with \sigs. For a variety of related studies in the
literature, using SPS, dynamics, and gravitational lensing,
similar results are found. Studies based solely on spectroscopic
line diagnostics agree on a Salpeter-like IMF at high \sigs, but
differ at low \sigs. Overall, we find that multiple independent
lines of evidence appear to be converging on a systematic
variation in the IMF, such that high-\sigs\ ETGs have an excess of
low-mass stars relative to spirals and low-\sigs\ ETGs. Robust
verification of super-Salpeter IMFs in the highest-\sigs\ galaxies
will require additional scrutiny of scatter and systematic
uncertainties. The implications for the distribution of DM are
still inconclusive.
\end{abstract}


\section{Introduction}\label{sec:intro}

The stellar Initial Mass Function (IMF) is fundamentally important to
understanding both stellar populations and galaxies.
The Milky Way (MW) IMF was originally characterized as
a power-law mass-distribution, $dN/dM \propto M^{-\alpha}$,
with $\alpha \sim 2.35$ \citep{Salpeter55}, and subsequently refined
to flatten at lower masses
($M \lsim 0.5 M_\odot$; \citealt{Kroupa01,Chabrier03}).

Whether or not the MW IMF describes stellar populations elsewhere
in the Universe cannot yet be said
through direct star counts.  There have been some
indirect observational hints of IMF variations, and ample
theoretical motivation for these, but no broadly convincing
evidence has emerged (cf.\ \citealt{Bastian+10}).

This situation has recently changed, with a flurry of studies of
early-type galaxies (ETGs) turning up
indirect evidence for systematic IMF variations. These studies use
models of stellar population synthesis (SPS) to fit
integrated-light data (broad-band colors and spectroscopic
features), and fall into two broad categories: ``pure'' SPS, and
``hybrid'' SPS+gravitating mass analyses.

The pure analyses rely on spectral lines that are
differentially sensitive to giant or dwarf stars.
These include the TiO feature at 6130~\AA, the Na~{\small I} doublet near 8190~\AA,
the Ca~{\small II} triplet near 8600~\AA, the Wing--Ford [Fe/H] band at 9915~\AA,
and the Ca~{\small I} line at 10345~\AA\
(e.g.,
\citealt{Cenarro+03,vDC10,Spiniello+12};
\citealt[hereafter CvD12]{Conroy_vanDokkum12b};
\citealt{Ferreras+12,Smith+12}).

The hybrid analyses assume an IMF and infer a stellar
mass-to-light ratio \Yst\ using a more conventional SPS approach
based on colors, and age- and metallicity-sensitive spectral
lines. Estimates of \Yst\ are also derived using dynamics or
gravitational lensing. Comparison of the independent results then
yields a revised IMF (e.g.,
\citealt{Cappellari+06,Cappellari+12,Cappellari+12_ATLAS3D_XX,FSB08,Tortora+09,Tortora+10lensing,Grillo_Cobat10,Grillo10,Treu+10,NRT10,Napolitano+11_PNS,Auger+10,Thomas+11,Spiniello+11,Dutton+12a,Dutton+12b,Sonnenfeld+12,Wegner+12,SPIDER-VI}).

One may characterize a revised IMF through its \Yst\ relative to a
MW-disk IMF, $\dimf\ \equiv \Yst/\Upsilon_{\star,{\rm MW}}$ (the
``mismatch parameter''), where for reference we adopt the Chabrier
IMF. Remarkably, almost all the above studies found ``heavy'' IMFs
($\dimf\ \gg 1$) for the most massive ETGs. Less massive ETGs, and
spiral galaxies, appear to have ``normal/light'' IMFs ($\dimf
\lsim 1$; e.g.,
\citealt{Bell_deJong01,Bershady+11,Suyu+12,Brewer+12}) and the
bulge components of spirals may also have a similar mass
dependence to ETGs (\citealt{deBlok+08};
\citealt[F+10]{Ferreras+10}; \citealt[D+12c]{Dutton+12c}).

These IMF findings are both potentially revolutionary, and highly controversial,
and demand further investigation.
In particular, with hybrid analyses there are lingering questions about
degeneracies associated with the distribution of non baryonic dark matter (DM).
The time is also ripe to inventory the results to date, and see
if the apparent emerging consensus holds up under quantitative, systematic comparison
-- which could provide pressing motivation for understanding
the physical origins of the trends.
Comparisons were made for some hybrid analyses
\citep{Thomas+11,Dutton+12b,Wegner+12}, but nor for the pure SPS work.

In this Letter we carry out such an inventory, while presenting
our own novel results for a large sample of ETGs for reference,
following the dynamical$+$SPS analyses of \citet[hereafter
T+12]{SPIDER-VI}. We focus on the trends in \dimf\ with central
stellar velocity dispersion, \sigs, and discuss some implications
for the central DM fraction. \sigs\ is widely considered as
crucially connected to galaxy evolution, and unlike \Yst, is
relatively independent of the bandpass, and of \dimf.

We describe our data and analysis methods in Section~\ref{sec:data}.
We present our results in Section~\ref{sec:results}
and make literature comparisons in Section~\ref{sec:lit}.
We summarize the conclusions and outlook in
Section~\ref{sec:conclusions}.

\section{Data and analysis methods}\label{sec:data}

We apply a combination of SPS and stellar dynamical models to a
sample of $\sim 4500$ giant ETGs, in the redshift range of
$z=0.05$--$0.1$, drawn from the SPIDER project
(\citealt{SPIDER-I}). Our data include optical$+$near-infrared
photometry [from the Sloan Digital Sky Survey (SDSS)  and the
UKIRT Infrared Deep Sky Survey-Large Area
Survey]\footnote{http://www.sdss.org, http://www.ukidss.org},
high-quality measurements of galactic structural parameters
(effective radius \Re\ and S\'ersic index $n$), and SDSS
central-aperture velocity dispersions $\sigAp$. The sample
galaxies are defined as bulge dominated systems with passive
spectra while late-type systems are efficiently removed through
the SDSS classification parameters based on the spectral type and
the fraction of light which is better described by a
\cite{deVauc48} profile (see T+12 for further details). The
structural parameters are measured using 2DPHOT
(\citealt{LaBarbera_08_2DPHOT}) and are found to be significantly
different from the SDSS estimates (\citealt{SPIDER-I}). The sample
is $95\%$ complete at a stellar mass of $\mst = 3 \times 10^{10}\,
\rm \Msun$, which corresponds to $\sigAp \sim 160 \, \rm km/s$.

The SPS-based \Yst\ values were derived by fitting \citet{BC03}
models to the multi-band photometry, assuming a Chabrier IMF.
These results have been shown to be consistent with independent
literature (e.g., MPA masses in \citealt{Dutton+12b}), while
possible systematics in stellar mass (\mst) estimates are
discussed in \cite{SPIDER-V} and T+12. Despite the well known
age-metallicity degeneracies in photometric data, these conspire
to keep the stellar mass-to-light ratio well constrained, with
scatter of $0.05-0.15$ dex. The agreement is also excellent when
our colour derived masses are compared with spectroscopic
estimates.  The largest systematical uncertainty in our analysis
comes from our ignorance on the IMF shape, which can produce
variations of the stellar M/L of a factor as large as $\sim 2$.
For this reason IMF is a key issue in stellar population analysis
and the central topic of this paper.

Our dynamical-mass estimates use spherical isotropic Jeans
equations fitted to the $\sigAp$ data. We also extend the T+12
analysis using two-component mass models: a S\'ersic-based stellar
distribution following the $K$-band light, and a standard DM
profile. For the latter we adopt a series of plausible assumptions
(cf.\ \citealt{Cappellari+12}), as the data do not allow us to
constrain both components simultaneously.

Our DM models are based on the \citet[hereafter NFW]{NFW96}
profile, with an adjustable degree of baryon-induced adiabatic
contraction (AC). For the virial mass and concentration ($\Mvir$,
$\cvir$), we adopt mean trends for a WMAP5 cosmology
\citep{Maccio+08}, for the \Mvir--\mst\ relation we used
\citet[M+10 hereafter]{Moster+10}. Each galaxy's mass model then
has one free parameter, \Yst, plus optional AC \citep[G+04
hereafter]{Gnedin+04}, providing our no-AC-NFW and AC-NFW
base-models.

We explore the sensitivity of our results to these assumptions by
doing the analyses with the following alternatives: a) AC recipes
of varying strengths (\citealt[B+86]{Blumenthal+86};
\citealt[A+10]{Abadi+10}); b) \Mvir\ with a fixed value for the
entire sample: $\Mvir = 10^{12} M_\odot$, $10^{13} M_\odot$, or
$10^{14} M_\odot$; c) WMAP3-based $\cvir$--$\Mvir$ relation
\citep{Maccio+08}; d) no DM is present; e) mild kinematic
anisotropy, with $\beta= -0.2$ or $+0.2$; f) $\cvir$--$\Mvir$
relation altered to mimic a warm dark matter (WDM) cosmology,
assuming different particle masses \citep[S+12a
hereafter]{Schneider+12}.

\begin{figure}
\centering
\includegraphics[width=0.35\textwidth,clip]{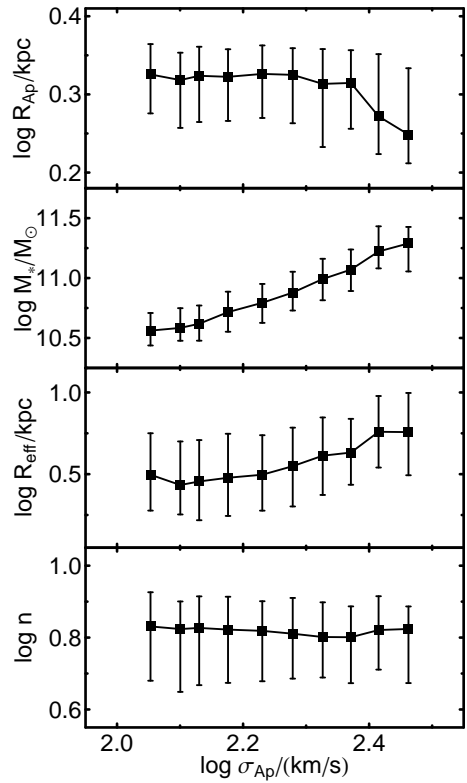}
\caption{Galaxy sample properties, binned by velocity dispersion:
physical aperture radius, SPS-based stellar mass (Chabrier IMF),
effective radius, and S\'ersic index. The points and error-bars
show the medians and 25--75 percentile scatter. }\label{fig:fig1}
\end{figure}

To study the mean trends of \Yst\ with velocity dispersion, we
construct ``average'' galaxies by dividing our sample into 10
\sigAp-bins, for which we compute median values of $M_\star$,
\Re\, $n$, \Rap, and \sigAp. We show these values, and their
25--75 percentile scatter, in Figure~\ref{fig:fig1}. For each
\sigAp-bin and a given DM model, we solve the radial Jeans
equation for the \Yst\ value which matches the observed \sigAp.

Our final analysis-products, for each galaxy bin and mass model,
will be the SPS-determined $\Upsilon_{\star, \rm MW}$, the
dynamically-determined \Yst, the inferred \dimf, and the inferred
central DM-fraction, $\fdm\equiv1-\Yst/\Ydyn$. For homogeneity, we
convert \sigAp\ to \sige\ (the value at \Re), using the
best-fitting relation in \cite{Cappellari+06} (also done for the
literature results later).

\vspace{0.5cm}

\section{Results: IMF and DM fraction trends}\label{sec:results}

\begin{figure*}[t]
\centering
\includegraphics[width=0.51\textwidth,clip]{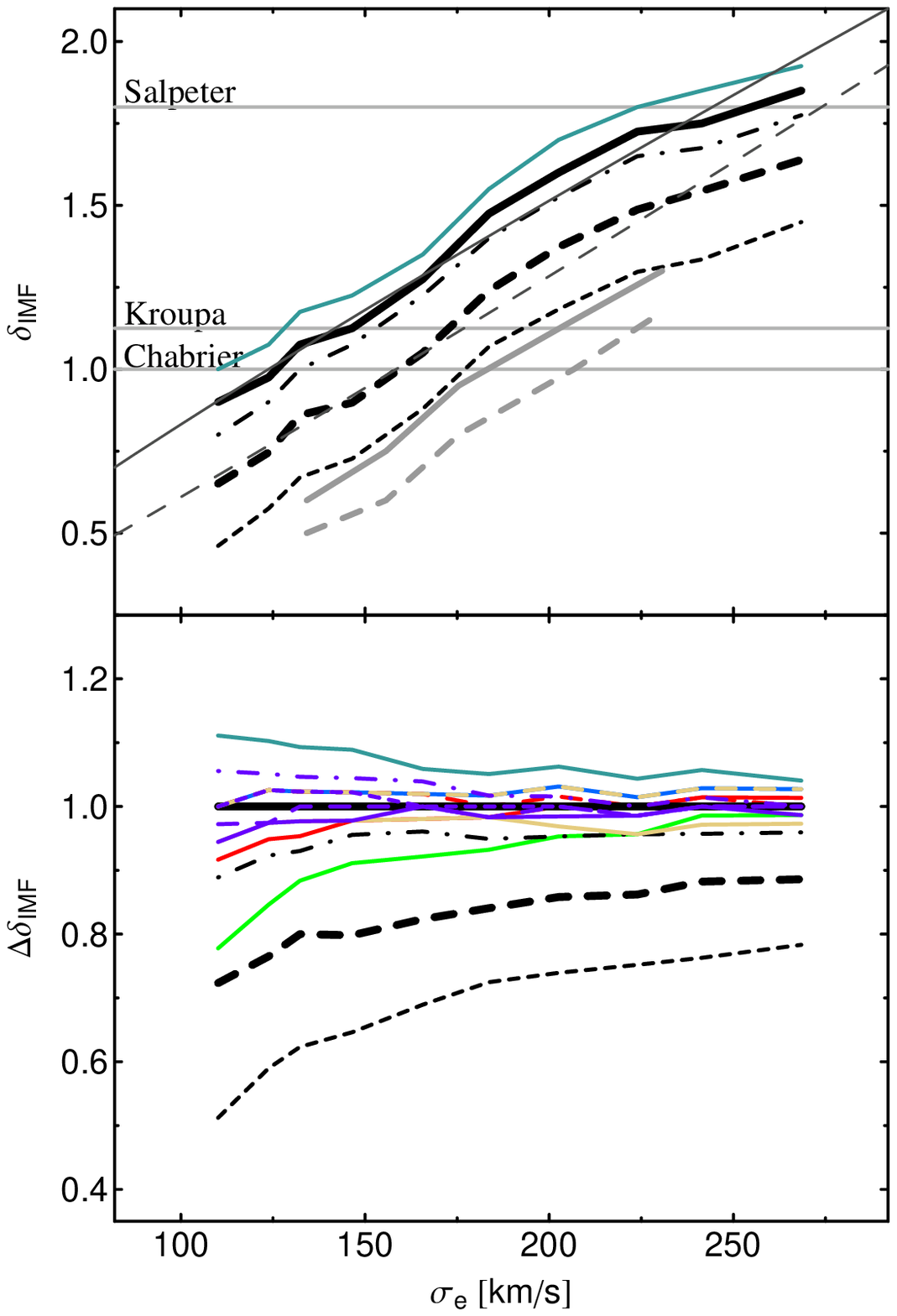}
\includegraphics[width=0.48\textwidth,clip]{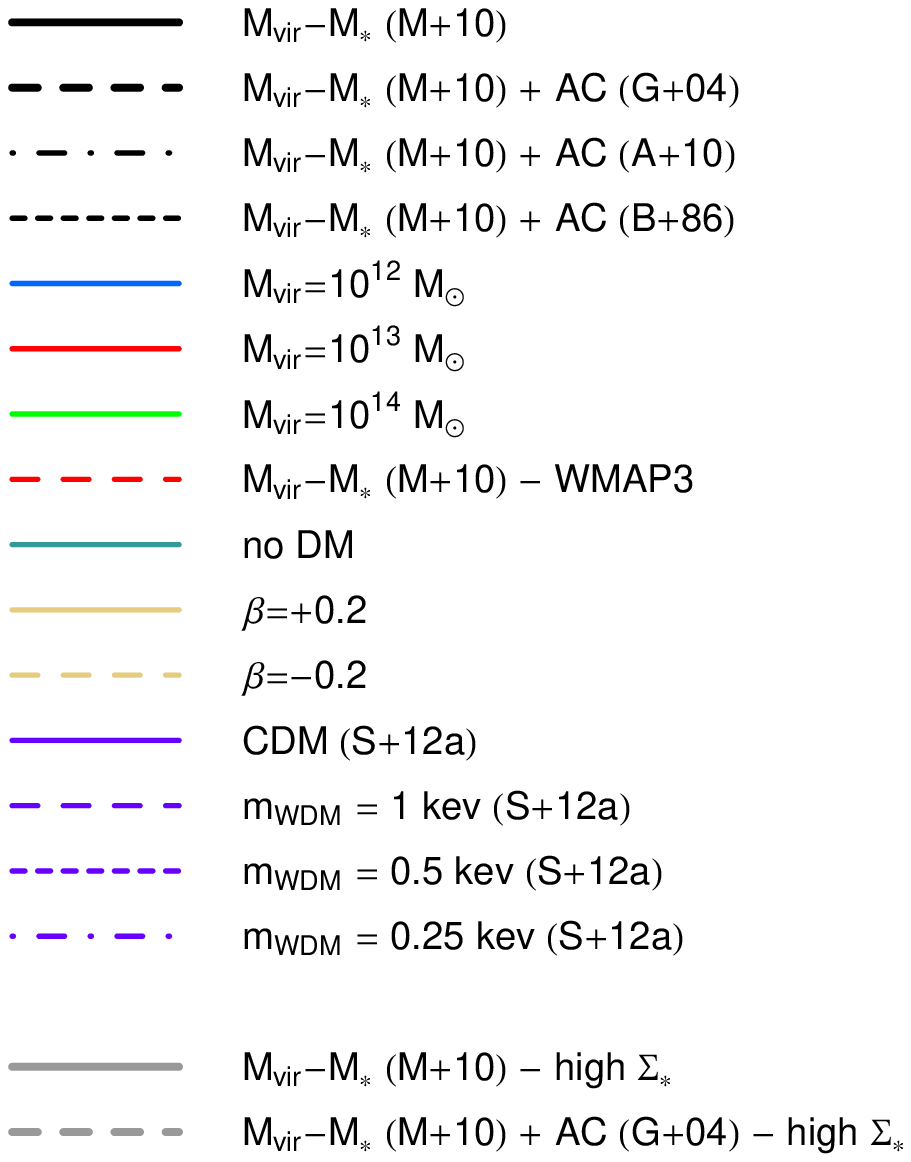}
\caption{IMF mismatch-parameter, $\delta_{\rm IMF}= \Yst /
\YstMW$, vs.\ velocity dispersion \sige, for the SPIDER sample.
{\it Top:} results for several different mass models. Horizontal
lines show reference values for Salpeter, Kroupa, Chabrier IMFs
(top-to-bottom). The best-fitted relations for the no-AC-NFW and
AC-NFW base-models are also shown as thin dark-grey lines. {\it
Bottom}: residuals relative to the fiducial no-AC-NFW result, for
an expanded series of mass models (see legend at right,
explanation in the main text, and reference list for
abbreviations). The overall normalization of \dimf\ depends on the
model assumptions, but a steep relation between \dimf\ and \sige\
emerges as a robust result. }\label{fig:spider}
\end{figure*}

Our main IMF results are shown
in Figure~\ref{fig:spider}. The two thick black curves
correspond to our standard no-AC-NFW (solid) and AC-NFW (long-dashed) models,
and the suite of alternative models are also plotted, as labeled.

It is clear that the overall normalization of \dimf\ is degenerate
with the adopted halo model, as DM can be traded against stellar
mass. This degeneracy is most severe when allowing for
uncertainties in the halo response to baryons (more so than with
the virial mass assumptions). However, for a given flavor of halo
model, there is always a strong correlation between \dimf\ and
\sige\ (the \dimf--\mst\ trend is weaker). We quantify this
correlation for each model with a log--log fit, reporting the
best-fit parameters in Table~\ref{tab:tab1} (also plotted in
Figure \ref{fig:spider} for the two reference models\footnote{We
note that a log--log fit although crude is a good approximation of
the trends in Fig. \ref{fig:spider}.}); the typical relation is
$\delta_{\rm IMF} \appropto \sigma_{\rm e}$.

\begin{table}
\centering \caption{Best-fit parameters for the relation $\log
\delta_{\rm IMF} = a + b \log \frac{\sige}{200 \, \rm km/s}$, for
model suite (see main text).}\label{tab:tab1}
\begin{tabular}{lcc} \hline
\rm Model & $a$ & $b$  \\
\hline
 \Mvir--\mst (M+10) & $0.18$ & $0.86$\\
 \Mvir--\mst (M+10) + AC (G+04) & $0.11$ & $1.07$\\
 \Mvir--\mst (M+10) + AC (A+10) & $0.16$ & $0.93$\\
 \Mvir--\mst (M+10) + AC (B+86) & $0.04$ & $1.29$\\
 $\Mvir = 10^{12} \rm \Msun$ & $0.19$ & $0.88$\\
 $\Mvir = 10^{13} \rm \Msun$ & $0.18$ & $0.97$\\
 $\Mvir = 10^{14} \rm \Msun$ & $0.16$ & $1.09$\\
 \Mvir--\mst (M+10) - WMAP3 & $0.18$ & $0.85$\\
 \rm no DM & $0.21$ & $0.79$\\
 $\beta = +0.2$ & $0.17$ & $0.87$\\
 $\beta = -0.2$ & $0.19$ & $0.88$\\
 CDM (S+11) & $0.17$ & $0.90$\\
 $m_{\rm WDM}= 1 \, \rm keV$ (S+11) & $0.18$ & $0.88$\\
 $m_{\rm WDM}= 0.5 \, \rm keV$ (S+11) & $0.18$ & $0.84$\\
 $m_{\rm WDM}= 0.25 \, \rm keV$ (S+11) & $0.19$ & $0.80$\\
 \hline
\end{tabular}
\end{table}

Other models of potential interest are low-density DM cores
\citep{Burkert95}, and alternatives to DM (e.g.,
\citealt{Milgrom83b}). These would imply similar \dimf--\sige\
slopes to our limiting no-DM model. The no-DM assumption produces
the uppermost curve in Fig. \ref{fig:spider}, and is qualitatively
consistent with a slightly {\it expanded} NFW model. The same
results, using the same dataset, but somewhat different mass
modeling assumptions, have been anticipated in \cite{SPIDER-VI}.

One could alter these slopes with additional model-tuning; e.g.,
with strong-AC at high-\sige\, and halo expansion at low-\sige,
one could completely flatten out the trend. On the other hand,
decreasing the AC at the high \sige\ and increasing the strength
of the AC toward lower \sige\, would make the trend even steeper.
We currently have no a priori motivation for either direction for
the AC variations, thus we cannot argue in favor of a IMF
universality for the former case nor for a very strong IMF non
universality in the latter one.

We next examine the overall IMF normalization, with the no-AC-NFW
and AC-NFW cases bracketing the most plausible range of models.
For reference, we show \dimf\ predictions for several standard
IMFs (Salpeter, Kroupa, Chabrier).  We note that at a fixed IMF,
age, and metallicity, the \Yst\ and \dimf\ values are uniquely
predicted, but the reverse is not true.  A given \dimf\ result can
imply multiple IMF solutions, particularly if one allows for
mass-functions more complicated than a pure power-law.

If we adopt a no-AC-NFW model, a MW-like IMF is implied for low-\sige\
galaxies, and a Salpeter IMF for high-\sige.
For AC-NFW, the low-\sige\ galaxies have sub-MW IMFs,
and the high-\sige\ ones have IMFs intermediate to Kroupa and Salpeter.

In all cases, extremely bottom-heavy IMFs (assumed single
power-law, $\alpha \gsim 2.6$) are ruled out, on average.  Even if
one assumed {\it no} DM, such IMFs would violate the dynamical
constraints on the overall mass-to-light ratio.

Although resolving the remaining IMF--DM degeneracy will require
more extensive analysis, we carry out a simple exercise to provide
initial clues, inspired by \citet{Dutton+12a}. As in that paper,
we select the galaxies with mean central stellar surface-densities
$\Sigma_\star >$~2500~$M_\odot$~pc$^{-2}$, and analyze them the
same as the full sample. The rationale is that such galaxies are
the most star-dominated, and the least sensitive to DM
uncertainties. The results are shown by the gray curves in
Figure~\ref{fig:spider}, where as expected, the model curves are
closer together. The implied \dimf\ normalization is fairly low --
similar to the full-sample AC results. We may then apply this IMF
result to the full sample if we assume no additional systematic
\dimf--$\Sigma_\star$ correlation (cf.\ \citealt{Schulz+10}).

Although this paper is primarily concerned with the IMF,
we briefly examine some implications for the central DM content.
Our main mass models are, by construction, fully consistent with
current expectations for $\Lambda$CDM halo profiles, while also agreeing with
the observations for a plausible IMF range ($\delta_{\rm IMF}\sim$~0.5--2.0).

\begin{figure}[t]
\centering
\includegraphics[width=0.49\textwidth,clip]{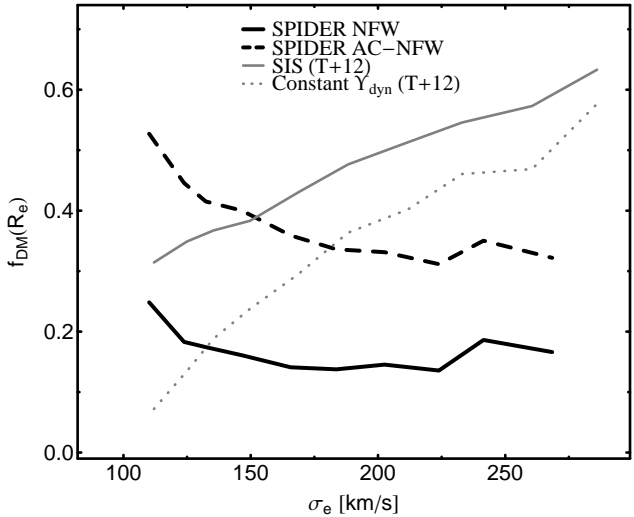}
\caption{Inferred DM fraction within $1 \, \Re$, vs.\ \sige, for
the SPIDER sample with different modeling assumptions. The solid
and dashed black curves are our default no-AC-NFW and AC-NFW
models. The gray curves are for fixed Chabrier IMF, with
isothermal and constant-\Ydyn\ models, bracketing the range of
non-parametric models (solid and dotted curves, respectively; see
T+12). With standard $\Lambda$CDM halos and a variable IMF, \fdm\
is constant or mildly decreasing with \sige; with a fixed IMF,
\fdm\ is strongly increasing (cf.\ \citealt{Thomas+11} figure
16).\vspace{1cm} }\label{fig:fdm}
\end{figure}

We show the implied \fdm\ within 1~\Re\ in Figure~\ref{fig:fdm}.
We find fairly universal values of $\fdm\sim0.2$ and $\sim0.5$
for the no-AC-NFW and AC-NFW models, respectively.
These results are not altered appreciably in the
alternative models explored above.
Note also that the high-$\Sigma_\star$ test above prefers the AC-NFW model.

The Figure also shows that if we adopted a constant IMF, then we
would infer a strong increase of \fdm\ with \sige. Such behavior
has been invoked as the driver for the ``tilt'' of the ETG
fundamental plane (see \citealt{Tortora+09} and references
therein), but now with renewed comparison to realistic DM halo
models, we find that the tilt is driven in part by the IMF. Put
differently, the observed \Ydyn--\sige\ relation is too steep to
explain through standard DM models, and requires an additional
factor.

\section{Literature comparisons}\label{sec:lit}

We now compare our SPIDER-based results with an inventory of other
literature results for ETGs in Figure~\ref{fig:lit} and late-type
galaxies in Figure~\ref{fig:lit2}.

\subsection{Early-type galaxies}

\begin{figure*}[t]
\centering
\includegraphics[trim= 6.5mm 0mm 5mm 0mm, width=0.70\textwidth,clip]{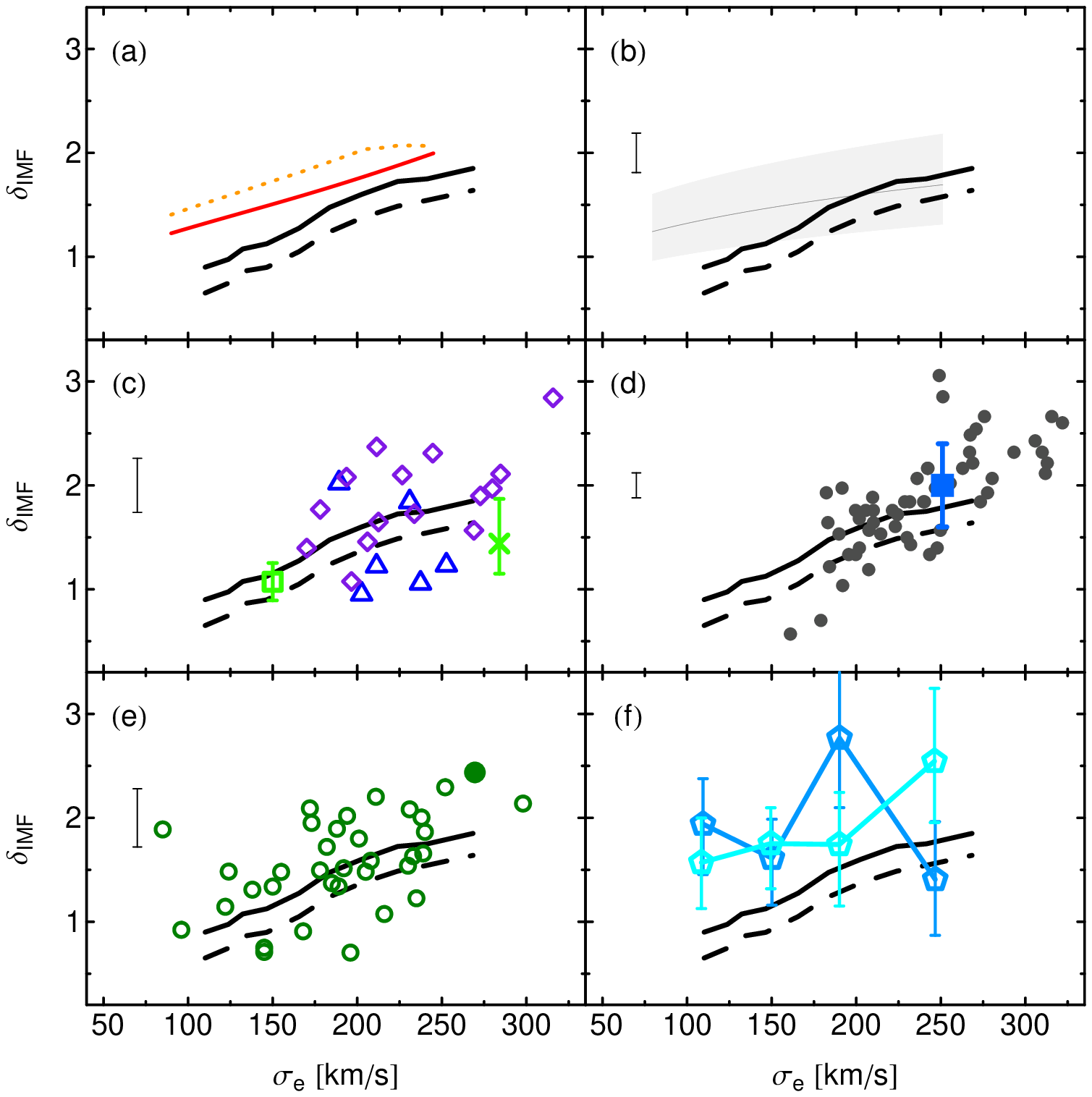}
\includegraphics[trim= 12.0mm -35mm 14mm 0mm, width=0.28\textwidth,clip]{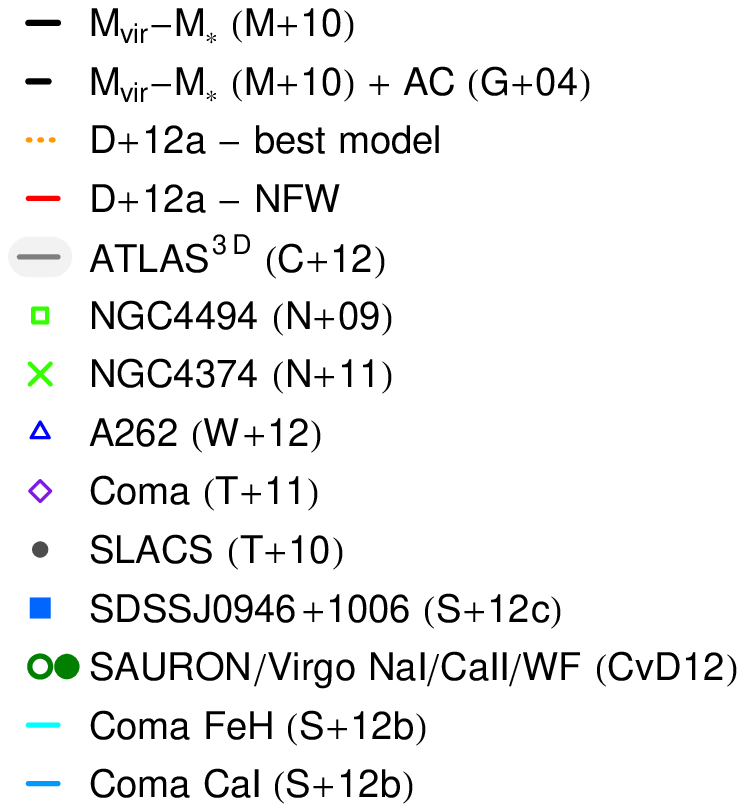}
\caption{IMF mismatch-parameter vs.\ stellar velocity dispersion,
comparing the SPIDER results (black curves) to various literature
studies. These include hybrid results from
\citet[D+12a]{Dutton+12b}, \citet[C+12]{Cappellari+12_ATLAS3D_XX},
\citet[W+12]{Wegner+12}, \citet[T+11]{Thomas+11} and
\citet[T+10]{Treu+10}, SPS ones from
\citet[CvD12]{Conroy_vanDokkum12b} and \citet[S+12b]{Smith+12},
and single-galaxy results from \citet[N+09]{Napolitano+09_PNS},
\citet[N+11]{Napolitano+11_PNS} and \citet[S+12c]{Sonnenfeld+12}.
In several panels, typical error-bars are shown to the left. The
filled CvD12 datapoint is from four galaxies' stacked spectra.
There is generally good consistency in the inferred IMF trends
from the different studies. }\label{fig:lit}
\end{figure*}

\begin{figure*}[t]
\centering
\includegraphics[width=0.5\textwidth,clip]{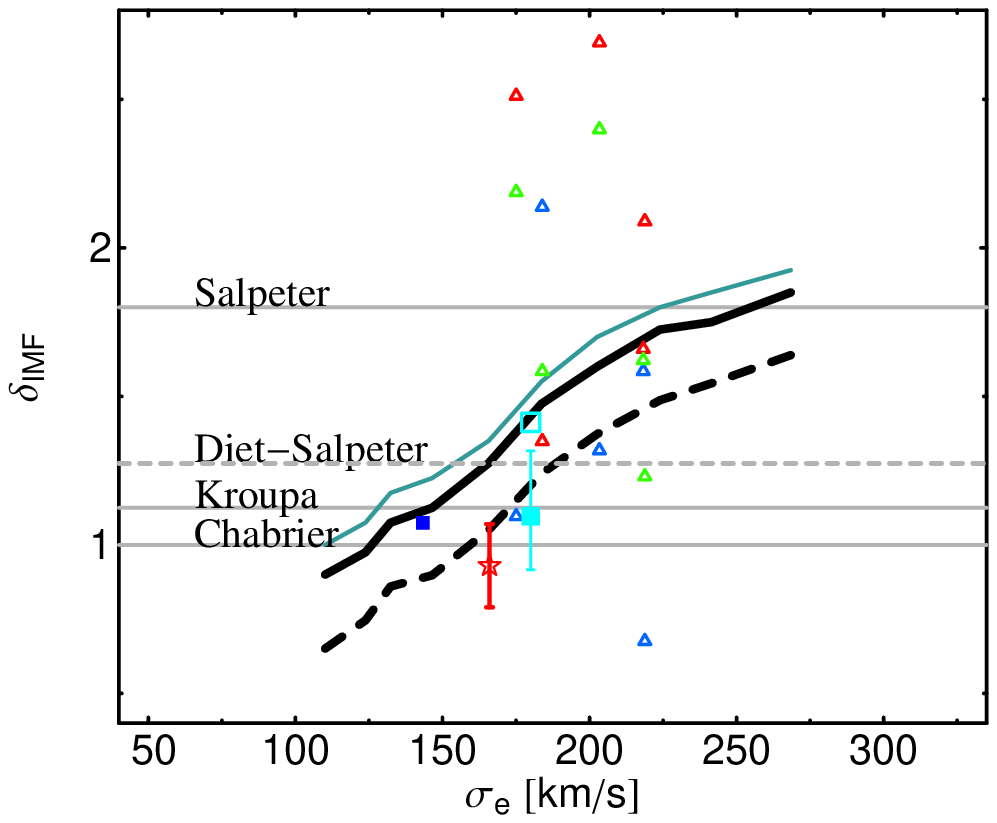}
\includegraphics[width=0.3\textwidth,clip]{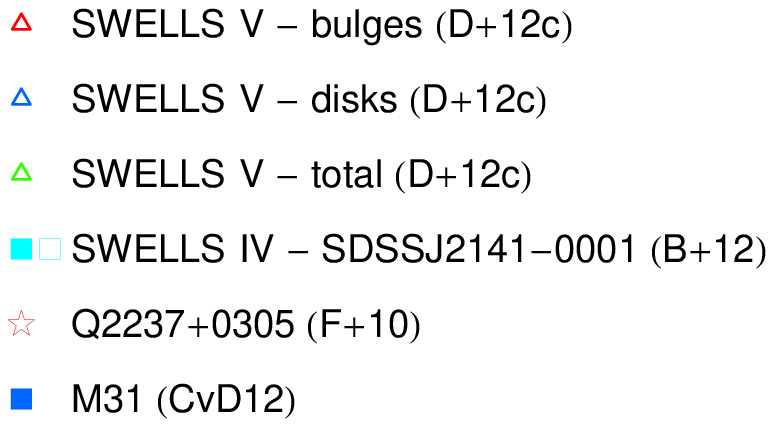}
\caption{IMF mismatch-parameter vs.\ \sigs, comparing the SPIDER
results to various literature studies on spiral galaxies. See
legend at right for the symbols (and legend in Fig.
\ref{fig:spider} for the curves). Horizontal solid lines show
reference values for Salpeter, Kroupa and Chabrier IMFs, while the
dashed one is for the diet-Salpeter from
\citet{Bell_deJong01}.}\label{fig:lit2}
\end{figure*}

We discuss the results for ETGs in Figure~\ref{fig:lit}, starting
with those studies that used similar hybrid approaches, comparing
SPS-mass estimates to total mass using dynamics or lensing. Rather
than exhaustively comparing all such results, we will focus on the
studies that explicitly derived \dimf\ for large samples of
low-$z$ ETGs.

The study most closely related to ours is \citet{Dutton+12b}, who
analyzed SDSS data for ETGs (colors and \sigs, with SPS$+$Jeans
modeling). Their sample was much larger but without the S\'ersic
models and near-infrared photometry from SPIDER. Their results for
a no-AC-NFW model are shown as a dotted orange curve in panel (a),
which is reassuringly similar to our no-AC-NFW result, with the
$\sim$~30\% residual difference in \dimf\ illustrating the level
of systematic uncertainties for a fixed dataset and method.
Although a conclusive answer on the origin of this discrepancy is
not available, we have found that shallower light profiles (as may
be equivalent to the combination of the $n=1$ and $n=4$ profiles
in \citealt{Dutton+12b}) produce larger \Yst. Finally, the solid
line shows their refined result from a multi-parameter fit: they
find DM halos that are slightly expanded, and consequently a
slightly heavier IMF.

In panel (b) we show results from the ATLAS$^{\rm 3D}$ survey of
nearby ETGs, using both spectroscopically-based SPS models and
detailed two-dimensional Jeans dynamical analyses \citep[where we
show their no-AC-NFW results]{Cappellari+12_ATLAS3D_XX}. As with
SPIDER, the ATLAS$^{\rm 3D}$ project found that the overall \dimf\
normalization was degenerate to the DM assumptions, but the trend
with \sige\ was robust. Direct comparison in Figure~\ref{fig:lit}
reveals that these results are consistent with SPIDER within the
errors, which is also the case for an earlier study of ETGs with
SAURON \citep{Cappellari+06} and for an analysis of high-$z$
galaxies  \citep{Cappellari+09}.

Panel (c) shows results from \citet{Thomas+11} and
\citet{Wegner+12}, who carried out spectroscopic SPS and detailed
orbit-modeling of ETGs found in clusters (including Coma), using a
variety of mass models (constant-\Ydyn, no-AC-NFW, and cored
halos). We see no systematic difference between the results
obtained with different mass models. A systematic offset in \dimf\
between the two studies is found, but overall, the results are
consistent with our no-AC-NFW results, both in amplitude and in
slope.

Panel (d) shows gravitational lensing results, primarily from the
SLACS survey of ETGs \citep{Treu+10}, using no-AC-NFW models as
well as color-based SPS models and stellar-dynamics constraints.
Their \dimf\ normalization agrees well overall with ours, but
their slope is somewhat steeper.\footnote{Not all of the analyses
of the SLACS data performed by non-SLACS teams agree about the IMF
conclusions. But we note that our own analysis in
\citet{Tortora+10lensing} agrees with the SLACS-team analysis.
Note also that the SPIDER and SLACS samples are selected on
luminosity and velocity dispersion, respectively.}

Our results, along with the four hybrid studies from the
literature, all suggest that the IMF of ETGs varies from MW-like
at low \sige\ to Salpeter-like at high \sige, modulo some
lingering uncertainties from the DM--IMF degeneracy.

In addition to the large-sample studies, we examine a few key
single-galaxy results.  These include extended kinematics data
with NFW-based modeling
\citep{Napolitano+09_PNS,Napolitano+11_PNS}, as shown in panel
(c), and the ``Jackpot'' double lens \citep{Sonnenfeld+12}, in
panel (d), and all look consistent with the trend from SPIDER.

Our culminating comparisons are with a completely different set of
results, based purely on modeling of
IMF-sensitive spectral lines (see Section~\ref{sec:intro}).
Most of these lines are susceptible to degeneracies with elemental
abundances (e.g., sodium or calcium), and we consider only those studies
that have directly accounted for such effects.

We first show in panel (e) the results from CvD12. They fitted
spectral features across a wide wavelength range, focusing on the
IMF indicators Na~{\small I}, Ca~{\small II}, and the Wing--Ford
band, while adopting a broken-power-law IMF form and fitting for
the relevant elemental abundances. Their inferred \dimf\ values
turn out to agree well with both the normalization and the trend
versus \sige\ from the SPIDER results (for the no-AC-NFW models in
particular)\footnote{CvD12 compared their \Yst\ results to {\it
total} dynamical values from SAURON in order to check that they
did not violate those constraints. However, they did not compare
to {\it decomposed} dynamical \Yst\ inferences for consistency as
we do here.}. Recalling that the high \dimf\ values from hybrid
studies could be due to either a bottom-heavy (extra dwarfs) {\it
or} a top-heavy (extra remnants) IMF, the CvD12 results agree with
only the first solution.

Note that the apertures probed here are different: $\Re/8$ for
CvD12, and 0.3--0.7~\Re\ for SPIDER (decreasing with \sigs). The
close agreement of the results on average thus implies that the
IMF does not vary spatially on these scales, or that the AC model
is the correct solution, and \dimf\ decreases with galactocentric
radius, which is plausible (e.g., \citealt{CVP86}; see also Fig.
13 of CvD12).

Panel (f) shows results from \citet{Smith+12}, who studied a large
sample of Coma-cluster ETGs. They used the same SPS models as
CvD12 to fit near-infrared spectroscopic line-indices, analyzing
the Wing--Ford band and the Ca~{\small I} line separately. The
former line was susceptible to uncertainties in the Na abundance,
but not the latter. We have converted their results to inferred
\dimf\ by straightforward interpolation between the Chabrier,
Salpeter, and $\alpha=3$ models in their figure~10.  The final
results are somewhat noisy and uncertain, but imply an overall
Salpeter-like normalization, and no obvious trend with \sige. This
agrees with all the aforementioned results at high-\sige, but not
at low-\sige.  It is possible that these Coma-cluster galaxies are
genuinely different. As environmental classification is part of
the SPIDER dataset (\citealt{SPIDER-II}; \citealt{SPIDER-III};
\citealt{SPIDER-VI}), we have investigated the impact of
environment on our results and found very little effect on \dimf.
But we also have very few cluster galaxies in our sample and a
definitive comparison with those Coma-cluster results is not
possible.

\citet{Spiniello+12} analyzed NaD, Na~{\small I}, and TiO$_2$
lines in SDSS spectra of ETGs with $\sigs > 200$~km~s$^{-1}$,
comparing to the same SPS models as CvD12. They inferred a
Salpeter-like IMF at low \sigs, which is consistent with SPIDER.
At high \sigs, they inferred $\alpha\sim3$, which would imply
$\dimf\sim 4$ and violate the total mass constraints both from
SPIDER and from the lens galaxy that these authors also studied.
This conflict suggests either further work is needed on the
line-index modeling, or the IMF shape deviates from a simple
power-law.

An ideal comparison with our SPIDER results would be the work of
\citet{Ferreras+12}, who analyzed Na~{\small I} and TiO
line-strengths from the SPIDER parent dataset. Although they did
not provide \dimf\ values that we could compare to our results,
their illustrative \Yst\ trends versus \sigs\ (at fixed
metallicity and age) are qualitatively similar.  They also
demonstrated that the \dimf\ inferences in their approach could
depend strongly on the detailed shape assumed for the IMF.

\subsection{Late-type galaxies}

It would also be fascinating to see whether or not bulges/disks
follow the same \dimf--\sigs\ trends as ETGs. Unfortunately,
although direct inferences from star counts are possible in the
Milky Way and nearby galaxies, the literature in the field is not
sufficient to investigate with accuracy IMF variations with mass
or \sigs, if any. However, it appears that a general consensus is
arising within the community, which points to Kroupa/Chabrier
type. In general, IMF looks similar in the field, dense massive
clusters and diffuse low density star-forming regions, with some
deviations observed in a handful of other regions (see
\citealt{Bastian+10} and \citealt{Kroupa+11} for a review of the
main results). Analysis of masses above $\sim 1 \, \Msun$ have
been performed in nearby galaxies (for instance the irregular LMC,
the dwarf SMC and the spiral M33), ruling out strong IMF
variations. Similarly, starburst galaxies and their embedded young
massive clusters imply no IMF variation and no influence of the
local environment.

Such analyses provide only limited information on a restricted
sample of galaxies. However, in the recent years, similarly to the
ETGs case, studies of variations with galaxy mass have been
accumulating. For example, \citet{Falcon+03} found that bulges of
spirals showed anti-correlations between \sigs\ and Ca~II
line-strengths, similar to ETGs, but they could not determine if
this was an IMF effect.

We have attempted to investigate further, using \dimf\ inferences
for bulges and disks from the literature (e.g.,
\citealt{deBlok+08}; F+10; \citealt[B+12]{Barnabe+12}; CvD12;
D+12c). This comparison is shown in Figure~\ref{fig:lit2}.

Using bulges from \cite{deBlok+08} we find that \dimf\ correlates
with stellar mass, giving hints of an ETG-like \dimf--\sigs\
correlation, but we also find the same for the disks, which would
be peculiar. Note that initial results on dynamical masses of
nearby spiral disks suggest $\dimf \sim 1$ (\citealt{Bershady+11,
Westfall+11}).

We show the results for a sample of five massive spiral galaxies
from \cite{Dutton+12c}, who found a) stellar mass carrying out a
photometric SPS and b) an independent IMF estimate, using strong
gravitational lensing and gas kinematics. The apparent scatter for
both bulges and disks is enormous, suggesting more work is needed
to address the systematic errors, and to understand any additional
trends with detailed morphology. However, the IMF normalization is
higher in the bulges than in the disks.

A similar analysis has been performed in B+12 (updating the
results in \citealt{Dutton+11b}) where gravitational lensing, gas
rotation curve and stellar kinematics for the lens galaxy SDSS
J2141-0001 are used. They found a \dimf\ which is fully consistent
with an intermediate-normalization IMF (in between a Salpeter and
a Chabrier IMF), but taking into account the expected cold gas
fraction (which had not been included in the fitting procedure)
they find lower \dimf, agreeing with a Chabrier/Kroupa IMF. A
bottom-light IMF is also found by F+10, who analyzed the strong
lensing features of the Einstein Cross (Q2237+0305), and
\cite{Conroy_vanDokkum12b} using spectral lines in the nuclear
bulge of M31.

\section{Discussion and conclusions}\label{sec:conclusions}

We have analyzed the dynamics and stellar populations of a large
sample of ETGs from the SPIDER project, and found compelling
evidence for heavier IMFs in the central regions of higher-\sigs\
galaxies. The IMF mismatch relative to Chabrier is $\dimf
\sim$~0.5--1.1 at $\sigs\sim$~125~km~s$^{-1}$, and $\sim$~1.2--1.8
at at $\sim$~250~km~s$^{-1}$. The \dimf--\sigs\ trend is robust to
a wide set of modeling assumptions, and accounts for much of the
tilt in the fundamental plane. The distribution of DM is
degenerate with the overall IMF normalization and difficult to
constrain, and therefore we have assumed that any halo contraction
is invariant with \sigs. Some ways to break the degeneracy between
IMF and halo contraction are to  a) analyze cases with extended
 velocity dispersion or X-ray emission profiles
(e.g., \citealt{Napolitano+09_PNS,Napolitano+11_PNS}), or b)
incorporate complementary data from strong/weak gravitational
lensing (e.g. \citealt{Auger+10}). However, we have argued that
the only way to preserve the IMF universality is to allow for halo
contraction at high-\sigs\, and halo {\it expansion} at low-\sigs.

We have performed the first general inventory of IMF results from
a variety of studies in the literature, using both pure and hybrid
techniques, and found that these generally agree well with our
SPIDER results. There is remarkably widespread agreement on a
Salpeter-like IMF for massive ETGs ($\sigs
\gsim$~200~km~s$^{-1}$). At lower \sigs, the data are still
somewhat limited and the different results have not yet converged,
but most of the studies point to MW-like IMFs. Agreement on a
super-Salpeter IMF in the most massive galaxies ($\sigs
\gsim$~275~km~s$^{-1}$) is also not yet universal.

These results appear to be fully compatible with underlying $\Lambda$CDM halos.
However, more detailed conclusions about halo contraction or expansion are
still elusive, and the data on galaxy centers do not clearly rule out alternative
DM models once the variable IMF is accounted for.

More work is clearly needed to understand the systematics in the
different analyses; to build up better statistics on a wide range
of galaxy types, environments, and redshifts; and to determine
which parameters correlate best with IMF variations (e.g.,
metallicity or starburst intensity; CvD12; \citealt{Smith+12}). It
may also be particularly helpful to venture beyond the centers of
galaxies, using data from a wide baseline in radius to help break
the IMF--DM degeneracies.

It appears we are nearing convergence on determining {\it what} the
basic components of galaxies are (distributions of stars and DM).
The next challenge will be to understand {\it why} these arrive at
their distributions.  What drives the power-spectrum in
cloud fragmentation and star formation?  How do baryonic processes
interact with and re-shape their surrounding DM halos?

\acknowledgments

We thank the referees for their constructive comments. We thank
Matt Auger, Michele Cappellari, Reinaldo de Carvalho, Aaron
Dutton, Francesco La Barbera, and Tommaso Treu for helpful
discussions. We also thank Charlie Conroy for helpful discussions
and for providing us with his updated results in tabular form. CT
was supported by the Swiss National Science Foundation and the
Forschungskredit at the University of Zurich. AJR was supported by
the National Science Foundation Grant AST-0909237.


\bibliographystyle{apj}

\begin{thebibliography}{61}
\expandafter\ifx\csname
natexlab\endcsname\relax\def\natexlab#1{#1}\fi

\bibitem[{{Abadi} {et~al.}(2010){Abadi}, {Navarro}, {Fardal}, {Babul}, \&
  {Steinmetz}}]{Abadi+10}
{Abadi}, M.~G., {Navarro}, J.~F., {Fardal}, M., {Babul}, A., \&
{Steinmetz}, M.
  2010, \mnras, 407, 435, (A+10)

\bibitem[{{Auger} {et~al.}(2010){Auger}, {Treu}, {Gavazzi}, {Bolton},
  {Koopmans}, \& {Marshall}}]{Auger+10}
{Auger}, M.~W., {Treu}, T., {Gavazzi}, R., {et~al.} 2010, \apjl,
721, L163

\bibitem[{{Barnab{\`e}} {et~al.}(2012){Barnab{\`e}}, {Dutton}, {Marshall},
  {Auger}, {Brewer}, {Treu}, {Bolton}, {Koo}, \& {Koopmans}}]{Barnabe+12}
{Barnab{\`e}}, M., {Dutton}, A.~A., {Marshall}, P.~J., {et~al.}
2012, \mnras,
  423, 1073, (B+12)

\bibitem[{{Bastian} {et~al.}(2010){Bastian}, {Covey}, \& {Meyer}}]{Bastian+10}
{Bastian}, N., {Covey}, K.~R., \& {Meyer}, M.~R. 2010, \araa, 48,
339

\bibitem[{{Bell} \& {de Jong}(2001)}]{Bell_deJong01}
{Bell}, E.~F., \& {de Jong}, R.~S. 2001, \apj, 550, 212

\bibitem[{{Bershady} {et~al.}(2011){Bershady}, {Martinsson}, {Verheijen},
  {Westfall}, {Andersen}, \& {Swaters}}]{Bershady+11}
{Bershady}, M.~A., {Martinsson}, T.~P.~K., {Verheijen}, M.~A.~W.,
{et~al.}
  2011, \apjl, 739, L47

\bibitem[{{Blumenthal} {et~al.}(1986){Blumenthal}, {Faber}, {Flores}, \&
  {Primack}}]{Blumenthal+86}
{Blumenthal}, G.~R., {Faber}, S.~M., {Flores}, R., \& {Primack},
J.~R. 1986,
  \apj, 301, 27, (B+86)

\bibitem[{{Brewer} {et~al.}(2012){Brewer}, {Dutton}, {Treu}, {Auger},
  {Marshall}, {Barnab{\`e}}, {Bolton}, {Koo}, \& {Koopmans}}]{Brewer+12}
{Brewer}, B.~J., {Dutton}, A.~A., {Treu}, T., {et~al.} 2012,
\mnras, 422, 3574

\bibitem[{{Bruzual} \& {Charlot}(2003)}]{BC03}
{Bruzual}, G., \& {Charlot}, S. 2003, \mnras, 344, 1000

\bibitem[{{Burkert}(1995)}]{Burkert95}
{Burkert}, A. 1995, \apjl, 447, L25

\bibitem[{{Cappellari} {et~al.}(2006){Cappellari}, {Bacon}, {Bureau}, {Damen},
  {Davies}, {de Zeeuw}, {Emsellem}, {Falc{\'o}n-Barroso}, {Krajnovi{\'c}},
  {Kuntschner}, {McDermid}, {Peletier}, {Sarzi}, {van den Bosch}, \& {van de
  Ven}}]{Cappellari+06}
{Cappellari}, M., {Bacon}, R., {Bureau}, M., {et~al.} 2006,
\mnras, 366, 1126,
  (C+06)

\bibitem[{{Cappellari} {et~al.}(2009){Cappellari}, {di Serego Alighieri},
  {Cimatti}, {Daddi}, {Renzini}, {Kurk}, {Cassata}, {Dickinson},
  {Franceschini}, {Mignoli}, {Pozzetti}, {Rodighiero}, {Rosati}, \&
  {Zamorani}}]{Cappellari+09}
{Cappellari}, M., {di Serego Alighieri}, S., {Cimatti}, A.,
{et~al.} 2009,
  \apjl, 704, L34

\bibitem[{{Cappellari} {et~al.}(2012{\natexlab{a}}){Cappellari}, {McDermid},
  {Alatalo}, {Blitz}, {Bois}, {Bournaud}, {Bureau}, {Crocker}, {Davies},
  {Davis}, {de Zeeuw}, {Duc}, {Emsellem}, {Khochfar}, {Krajnovi{\'c}},
  {Kuntschner}, {Lablanche}, {Morganti}, {Naab}, {Oosterloo}, {Sarzi}, {Scott},
  {Serra}, {Weijmans}, \& {Young}}]{Cappellari+12}
{Cappellari}, M., {McDermid}, R.~M., {Alatalo}, K., {et~al.}
  2012{\natexlab{a}}, \nat, 484, 485

\bibitem[{{Cappellari} {et~al.}(2012{\natexlab{b}}){Cappellari}, {McDermid},
  {Alatalo}, {Blitz}, {Bois}, {Bournaud}, {Bureau}, {Crocker}, {Davies},
  {Davis}, {de Zeeuw}, {Duc}, {Khochfar}, {Krajnovic}, {Kuntschner},
  {Morganti}, {Naab}, {Oosterloo}, {Sarzi}, {Scott}, {Serra}, {Weijmans}, \&
  {Young}}]{Cappellari+12_ATLAS3D_XX}
---. 2012{\natexlab{b}}, \mnras, submitted, arXiv:1208.3523, (C+12)

\bibitem[{{Carter} {et~al.}(1986){Carter}, {Visvanathan}, \& {Pickles}}]{CVP86}
{Carter}, D., {Visvanathan}, N., \& {Pickles}, A.~J. 1986, \apj,
311, 637

\bibitem[{{Cenarro} {et~al.}(2003){Cenarro}, {Gorgas}, {Vazdekis}, {Cardiel},
  \& {Peletier}}]{Cenarro+03}
{Cenarro}, A.~J., {Gorgas}, J., {Vazdekis}, A., {Cardiel}, N., \&
{Peletier},
  R.~F. 2003, \mnras, 339, L12

\bibitem[{{Chabrier}(2003)}]{Chabrier03}
{Chabrier}, G. 2003, \pasp, 115, 763

\bibitem[{{Conroy} \& {van Dokkum}(2012)}]{Conroy_vanDokkum12b}
{Conroy}, C., \& {van Dokkum}, P.~G. 2012, \apj, 760, 71

\bibitem[{{de Blok} {et~al.}(2008){de Blok}, {Walter}, {Brinks},
  {Trachternach}, {Oh}, \& {Kennicutt}}]{deBlok+08}
{de Blok}, W.~J.~G., {Walter}, F., {Brinks}, E., {et~al.} 2008,
\aj, 136, 2648

\bibitem[{{de Vaucouleurs}(1948)}]{deVauc48}
{de Vaucouleurs}, G. 1948, Annales d'Astrophysique, 11, 247

\bibitem[{{Dutton} {et~al.}(2012{\natexlab{a}}){Dutton}, {Maccio'}, {Mendel},
  \& {Simard}}]{Dutton+12b}
{Dutton}, A.~A., {Maccio'}, A.~V., {Mendel}, J.~T., \& {Simard},
L.
  2012{\natexlab{a}}, \mnras, submitted, arXiv:1204.2825, (D+12a)

\bibitem[{{Dutton} {et~al.}(2012{\natexlab{b}}){Dutton}, {Mendel}, \&
  {Simard}}]{Dutton+12a}
{Dutton}, A.~A., {Mendel}, J.~T., \& {Simard}, L.
2012{\natexlab{b}}, \mnras,
  422, L33

\bibitem[{{Dutton} {et~al.}(2011){Dutton}, {Brewer}, {Marshall}, {Auger},
  {Treu}, {Koo}, {Bolton}, {Holden}, \& {Koopmans}}]{Dutton+11b}
{Dutton}, A.~A., {Brewer}, B.~J., {Marshall}, P.~J., {et~al.}
2011, \mnras,
  417, 1621

\bibitem[{{Dutton} {et~al.}(2012{\natexlab{c}}){Dutton}, {Treu}, {Brewer},
  {Marshall}, {Auger}, {Barnabe}, {Koo}, {Bolton}, \& {Koopmans}}]{Dutton+12c}
{Dutton}, A.~A., {Treu}, T., {Brewer}, B.~J., {et~al.}
2012{\natexlab{c}},
  \mnras, in press, arXiv:1206.4310, (D+12c)

\bibitem[{{Falc{\'o}n-Barroso} {et~al.}(2003){Falc{\'o}n-Barroso}, {Peletier},
  {Vazdekis}, \& {Balcells}}]{Falcon+03}
{Falc{\'o}n-Barroso}, J., {Peletier}, R.~F., {Vazdekis}, A., \&
{Balcells}, M.
  2003, \apjl, 588, L17

\bibitem[{{Ferreras} {et~al.}(2012){Ferreras}, {La Barbera}, {de Carvalho}, {de
  la Rosa}, {Vazdekis}, {Falcon-Barroso}, \& {Ricciardelli}}]{Ferreras+12}
{Ferreras}, I., {La Barbera}, F., {de Carvalho}, R.~R., {et~al.}
2012, \mnras,
  in press, arXiv:1204.3823

\bibitem[{{Ferreras} {et~al.}(2008){Ferreras}, {Saha}, \& {Burles}}]{FSB08}
{Ferreras}, I., {Saha}, P., \& {Burles}, S. 2008, \mnras, 383, 857

\bibitem[{{Ferreras} {et~al.}(2010){Ferreras}, {Saha}, {Leier}, {Courbin}, \&
  {Falco}}]{Ferreras+10}
{Ferreras}, I., {Saha}, P., {Leier}, D., {Courbin}, F., \&
{Falco}, E.~E. 2010,
  \mnras, 409, L30, (F+10)

\bibitem[{{Gnedin} {et~al.}(2004){Gnedin}, {Kravtsov}, {Klypin}, \&
  {Nagai}}]{Gnedin+04}
{Gnedin}, O.~Y., {Kravtsov}, A.~V., {Klypin}, A.~A., \& {Nagai},
D. 2004, \apj,
  616, 16, (G+04)

\bibitem[{{Grillo}(2010)}]{Grillo10}
{Grillo}, C. 2010, \apj, 722, 779

\bibitem[{{Grillo} \& {Gobat}(2010)}]{Grillo_Cobat10}
{Grillo}, C., \& {Gobat}, R. 2010, \mnras, 402, L67

\bibitem[{{Kroupa}(2001)}]{Kroupa01}
{Kroupa}, P. 2001, \mnras, 322, 231

\bibitem[{{Kroupa} {et~al.}(2011){Kroupa}, {Weidner}, {Pflamm-Altenburg},
  {Thies}, {Dabringhausen}, {Marks}, \& {Maschberger}}]{Kroupa+11}
{Kroupa}, P., {Weidner}, C., {Pflamm-Altenburg}, J., {et~al.}
2011, ArXiv
  e-prints

\bibitem[{{La Barbera} {et~al.}(2010{\natexlab{a}}){La Barbera}, {de Carvalho},
  {de La Rosa}, \& {Lopes}}]{SPIDER-II}
{La Barbera}, F., {de Carvalho}, R.~R., {de La Rosa}, I.~G., \&
{Lopes},
  P.~A.~A. 2010{\natexlab{a}}, \mnras, 408, 1335

\bibitem[{{La Barbera} {et~al.}(2010{\natexlab{b}}){La Barbera}, {de Carvalho},
  {de La Rosa}, {Lopes}, {Kohl-Moreira}, \& {Capelato}}]{SPIDER-I}
{La Barbera}, F., {de Carvalho}, R.~R., {de La Rosa}, I.~G.,
{et~al.}
  2010{\natexlab{b}}, \mnras, 408, 1313

\bibitem[{{La Barbera} {et~al.}(2008){La Barbera}, {de Carvalho},
  {Kohl-Moreira}, {Gal}, {Soares-Santos}, {Capaccioli}, {Santos}, \&
  {Sant'anna}}]{LaBarbera_08_2DPHOT}
{La Barbera}, F., {de Carvalho}, R.~R., {Kohl-Moreira}, J.~L.,
{et~al.} 2008,
  \pasp, 120, 681

\bibitem[{{La Barbera} {et~al.}(2010{\natexlab{c}}){La Barbera}, {Lopes}, {de
  Carvalho}, {de La Rosa}, \& {Berlind}}]{SPIDER-III}
{La Barbera}, F., {Lopes}, P.~A.~A., {de Carvalho}, R.~R., {de La
Rosa}, I.~G.,
  \& {Berlind}, A.~A. 2010{\natexlab{c}}, \mnras, 408, 1361

\bibitem[{{Macci{\`o}} {et~al.}(2008){Macci{\`o}}, {Dutton}, \& {van den
  Bosch}}]{Maccio+08}
{Macci{\`o}}, A.~V., {Dutton}, A.~A., \& {van den Bosch}, F.~C.
2008, \mnras,
  391, 1940

\bibitem[{{Milgrom}(1983)}]{Milgrom83b}
{Milgrom}, M. 1983, \apj, 270, 371

\bibitem[{{Moster} {et~al.}(2010){Moster}, {Somerville}, {Maulbetsch}, {van den
  Bosch}, {Macci{\`o}}, {Naab}, \& {Oser}}]{Moster+10}
{Moster}, B.~P., {Somerville}, R.~S., {Maulbetsch}, C., {et~al.}
2010, \apj,
  710, 903, (M+10)

\bibitem[{{Napolitano} {et~al.}(2010){Napolitano}, {Romanowsky}, \&
  {Tortora}}]{NRT10}
{Napolitano}, N.~R., {Romanowsky}, A.~J., \& {Tortora}, C. 2010,
\mnras, 405,
  2351

\bibitem[{{Napolitano} {et~al.}(2009){Napolitano}, {Romanowsky}, {Coccato},
  {Capaccioli}, {Douglas}, {Noordermeer}, {Gerhard}, {Arnaboldi}, {de Lorenzi},
  {Kuijken}, {Merrifield}, {O'Sullivan}, {Cortesi}, {Das}, \&
  {Freeman}}]{Napolitano+09_PNS}
{Napolitano}, N.~R., {Romanowsky}, A.~J., {Coccato}, L., {et~al.}
2009, \mnras,
  393, 329, (N+09)

\bibitem[{{Napolitano} {et~al.}(2011){Napolitano}, {Romanowsky}, {Capaccioli},
  {Douglas}, {Arnaboldi}, {Coccato}, {Gerhard}, {Kuijken}, {Merrifield},
  {Bamford}, {Cortesi}, {Das}, \& {Freeman}}]{Napolitano+11_PNS}
{Napolitano}, N.~R., {Romanowsky}, A.~J., {Capaccioli}, M.,
{et~al.} 2011,
  \mnras, 411, 2035, (N+11)

\bibitem[{{Navarro} {et~al.}(1996){Navarro}, {Frenk}, \& {White}}]{NFW96}
{Navarro}, J.~F., {Frenk}, C.~S., \& {White}, S.~D.~M. 1996, \apj,
462, 563

\bibitem[{{Salpeter}(1955)}]{Salpeter55}
{Salpeter}, E.~E. 1955, \apj, 121, 161

\bibitem[{{Schneider} {et~al.}(2012){Schneider}, {Smith}, {Macci{\`o}}, \&
  {Moore}}]{Schneider+12}
{Schneider}, A., {Smith}, R.~E., {Macci{\`o}}, A.~V., \& {Moore},
B. 2012,
  \mnras, 424, 684, (S+12a)

\bibitem[{{Schulz} {et~al.}(2010){Schulz}, {Mandelbaum}, \&
  {Padmanabhan}}]{Schulz+10}
{Schulz}, A.~E., {Mandelbaum}, R., \& {Padmanabhan}, N. 2010,
\mnras, 408, 1463

\bibitem[{{Smith} {et~al.}(2012){Smith}, {Lucey}, \& {Carter}}]{Smith+12}
{Smith}, R.~J., {Lucey}, J.~R., \& {Carter}, D. 2012, \mnras, 426,
(S+12b)

\bibitem[{{Sonnenfeld} {et~al.}(2012){Sonnenfeld}, {Treu}, {Gavazzi},
  {Marshall}, {Auger}, {Suyu}, {Koopmans}, \& {Bolton}}]{Sonnenfeld+12}
{Sonnenfeld}, A., {Treu}, T., {Gavazzi}, R., {et~al.} 2012, \apj,
752, 163,
  (S+12c)

\bibitem[{{Spiniello} {et~al.}(2011){Spiniello}, {Koopmans}, {Trager},
  {Czoske}, \& {Treu}}]{Spiniello+11}
{Spiniello}, C., {Koopmans}, L.~V.~E., {Trager}, S.~C., {Czoske},
O., \&
  {Treu}, T. 2011, \mnras, 417, 3000

\bibitem[{{Spiniello} {et~al.}(2012){Spiniello}, {Trager}, {Koopmans}, \&
  {Chen}}]{Spiniello+12}
{Spiniello}, C., {Trager}, S.~C., {Koopmans}, L.~V.~E., \& {Chen},
Y.~P. 2012,
  \apjl, 753, L32

\bibitem[{{Suyu} {et~al.}(2012){Suyu}, {Hensel}, {McKean}, {Fassnacht}, {Treu},
  {Halkola}, {Norbury}, {Jackson}, {Schneider}, {Thompson}, {Auger},
  {Koopmans}, \& {Matthews}}]{Suyu+12}
{Suyu}, S.~H., {Hensel}, S.~W., {McKean}, J.~P., {et~al.} 2012,
\apj, 750, 10

\bibitem[{{Swindle} {et~al.}(2011){Swindle}, {Gal}, {La Barbera}, \& {de
  Carvalho}}]{SPIDER-V}
{Swindle}, R., {Gal}, R.~R., {La Barbera}, F., \& {de Carvalho},
R.~R. 2011,
  \aj, 142, 118

\bibitem[{{Thomas} {et~al.}(2011){Thomas}, {Saglia}, {Bender}, {Thomas},
  {Gebhardt}, {Magorrian}, {Corsini}, {Wegner}, \& {Seitz}}]{Thomas+11}
{Thomas}, J., {Saglia}, R.~P., {Bender}, R., {et~al.} 2011,
\mnras, 415, 545,
  (T+11)

\bibitem[{{Tortora} {et~al.}(2012){Tortora}, {La Barbera}, {Napolitano}, {de
  Carvalho}, \& {Romanowsky}}]{SPIDER-VI}
{Tortora}, C., {La Barbera}, F., {Napolitano}, N.~R., {de
Carvalho}, R.~R., \&
  {Romanowsky}, A.~J. 2012, \mnras, 425, 577

\bibitem[{{Tortora} {et~al.}(2009){Tortora}, {Napolitano}, {Romanowsky},
  {Capaccioli}, \& {Covone}}]{Tortora+09}
{Tortora}, C., {Napolitano}, N.~R., {Romanowsky}, A.~J.,
{Capaccioli}, M., \&
  {Covone}, G. 2009, \mnras, 396, 1132

\bibitem[{{Tortora} {et~al.}(2010){Tortora}, {Napolitano}, {Romanowsky}, \&
  {Jetzer}}]{Tortora+10lensing}
{Tortora}, C., {Napolitano}, N.~R., {Romanowsky}, A.~J., \&
{Jetzer}, P. 2010,
  \apjl, 721, L1

\bibitem[{{Treu} {et~al.}(2010){Treu}, {Auger}, {Koopmans}, {Gavazzi},
  {Marshall}, \& {Bolton}}]{Treu+10}
{Treu}, T., {Auger}, M.~W., {Koopmans}, L.~V.~E., {et~al.} 2010,
\apj, 709,
  1195, (T+10)

\bibitem[{{van Dokkum} \& {Conroy}(2010)}]{vDC10}
{van Dokkum}, P.~G., \& {Conroy}, C. 2010, \nat, 468, 940

\bibitem[{{Wegner} {et~al.}(2012){Wegner}, {Corsini}, {Thomas}, {Saglia},
  {Bender}, \& {Pu}}]{Wegner+12}
{Wegner}, G.~A., {Corsini}, E.~M., {Thomas}, J., {et~al.} 2012,
\aj, 144, 78,
  (W+12)

\bibitem[{{Westfall} {et~al.}(2011){Westfall}, {Bershady}, {Verheijen},
  {Andersen}, {Martinsson}, {Swaters}, \& {Schechtman-Rook}}]{Westfall+11}
{Westfall}, K.~B., {Bershady}, M.~A., {Verheijen}, M.~A.~W.,
{et~al.} 2011,
  \apj, 742, 18

\end{thebibliography}

\end{document}